# TRANSIENT ATTACKS AGAINST THE VMG-KLJN SECURE KEY EXCHANGER

SHAHRIAR FERDOUS[1,+], LASZLO B. KISH[1,2]

[1]*Department of Electrical and Computer Engineering, Texas A&M University, College Station, TX 77841-3128, USA*
*ferdous.shahriar@tamu.edu , laszlokish@tamu.edu*

[2]*Óbuda University, Budapest, Bécsi út 96/B, Budapest, H-1034, Hungary*

**Abstract:** The security vulnerability of the Vadai, Mingesz, and Gingl (VMG) Kirchhoff-Law-Johnson-Noise (KLJN) key exchanger, as presented in the publication "Nature, Science Report 5 (2015) 13653," has been exposed to transient attacks. Recently an effective defense protocol was introduced (Appl. Phys. Lett. 122 (2023) 143503) to counteract mean-square voltage-based (or mean-square current-based) transient attacks targeted at the ideal KLJN framework.

In the present study, this same mitigation methodology has been employed to fortify the security of the VMG-KLJN key exchanger. It is worth noting that the protective measures need to be separately implemented for the HL and LH scenarios. This conceptual framework is corroborated through computer simulations, demonstrating that the application of this defensive technique substantially mitigates information leakage to a point of insignificance.

## 1. Introduction

Data encryption plays a pivotal role in ensuring the confidentiality and integrity of sensitive information [1-100]. In the realm of secure communications reliant on symmetric key-based mechanisms, the establishment of a secure key exchange is of paramount importance. Achieving information-theoretic (unconditional) security [1-7] in this context is made possible through two distinctive approaches: Quantum Key Distribution (QKD) [8-44] and the Kirchhoff-Law-Johnson-Noise (KLJN) secure key exchanger [3-7,45-100]. QKD leverages the principles of the quantum no-cloning theorem, a fundamental concept in quantum mechanics, to deliver unassailable security in key exchange processes [8-44]. In contrast, the KLJN scheme derives its security from classical statistical physics, drawing

---

[+] Corresponding Author

upon key principles such as the Fluctuation Dissipation Theorem [3], Kirchhoff's Law, and the characteristics of Gaussian stochastic processes [63].

In a recent publication [100], a highly effective defense protocol was introduced to counteract mean-square voltage (or current) based transient attacks [47,53] targeting the ideal Kirchhoff-Law-Johnson-Noise (KLJN) scheme. In the current study, we demonstrate that the VMG (Vadai-Mingesz-Gingl) variant of the KLJN key exchanger, as originally devised by Vadai, Mingesz, and Gingl [49] (refer to Section 1.2 for details), is equally susceptible to transient attacks as the ideal KLJN scheme. Furthermore, we establish that the same protective protocol [100] can be employed to enhance security in both scenarios.

*1.1. The KLJN key exchanger*

The KLJN system, as detailed in references [3-7], comprises a wireline connecting two communicating parties, denoted as Alice (A) and Bob (B), as depicted in Figure 1. At the heart of this key exchange mechanism are a pair of identical resistors, denoted as $R_H$ and $R_L$ (where $R_H > R_L$), along with two switches positioned at the ends of the communication line. During each bit exchange period (BEP), Alice and Bob independently and randomly select one of the resistors and establish a connection, thus forming a closed loop for the entire BEP. The key exchanger has the capability to function using either the intrinsic thermal noise generated by the resistors or through the introduction of proper external Gaussian voltage noise generators [50]. Importantly, the ideal KLJN system requires thermal equilibrium, that is, the uniformity of temperature (T) throughout the entire system, as described in references [83-84,86]. Moreover, the Gaussianity of the noises is essential for perfect security in the KLJN scheme, see the general proof in [60,63].

During the Bit Exchange Period (BEP), if Alice and Bob opt for different resistors, the ideal Kirchhoff-Law-Johnson-Noise (KLJN) system, characterized by its absence of delays or transients [3], ensures security, as outlined below. While passive eavesdropping enables Eve to measure the wire's voltage and current, allowing her to ascertain the resultant resistance within the loop [3,5], it remains beyond her capacity to discern which side holds the resistor $R_H$ and which side contains the resistor $R_L$. Consequently, her information entropy regarding the value of the key bit stands at 1 bit, rendering the bit secure [50-51,100].

The scenarios denoted as HL (where Alice connects to $R_H$ and Bob connects $R_L$), and LH (where Alice connects to $R_L$ and Bob connects $R_H$) correspond to the two possible values of the mutually shared secure bit [50-51,100]. For these situations, Alice and Bob establish a public agreement regarding the interpretation of the bit values, designating them as either 0 or 1 for the HL and LH configurations, or vice versa, respectively [50-51,100]. These configurations collectively facilitate a secure bit exchange process.

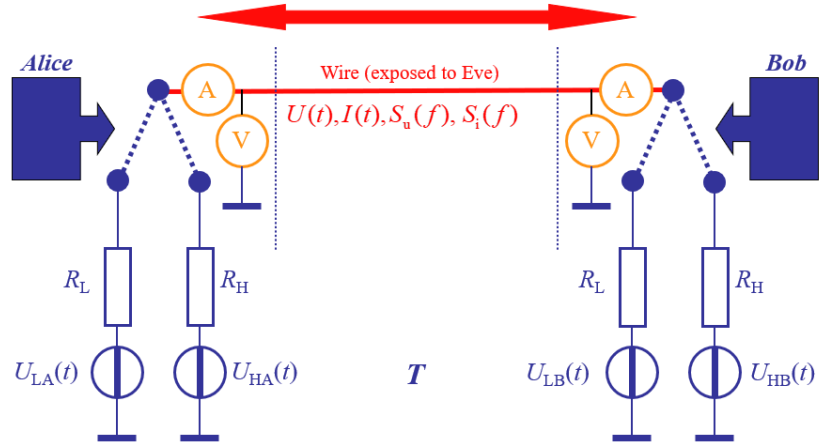

**Fig. 1.** The core of the KLJN secure key exchanger system consists of a wire line connection between the two communicating parties Alice and Bob [50-51,100]. The voltage $U(t)$, the current $I(t)$, and their corresponding spectra $S_u(f)$ and $S_i(f)$, respectively, are measurable by Alice, Bob and Eve. In the private space of Alice and Bob, the voltage generators $U_{HA}(t)$, $U_{LA}(t)$, $U_{HB}(t)$ and $U_{LB}(t)$ represent the independent thermal (Johnson-Nyquist) noises of the resistors (optionally external Gaussian noise generators for higher noise temperatures). The homogeneous temperature $T$ in the system guarantees that the HL (Alice $R_H$, Bob $R_L$) and LH (Alice $R_L$, Bob $R_H$) resistor connections provide identical mean-square voltage and current and the related spectra in the wire [50-51,100], and that the net power flow is zero between Alice and Bob.

Conversely, if, during the Bit Exchange Period (BEP), both parties opt for the same resistance value, resulting in either HH (where both Alice and Bob connect to $R_H$) or LL (where both Alice and Bob connect to $R_L$), the security of the system is compromised, as depicted in Figure 2. In these scenarios, Eve gains the capability to deduce the specific resistor configuration [50-51,100]. Consequently, the HH and LL bit configurations are deemed insecure and are rejected. The sole secure configurations encompass the HL ($R_H$, $R_L$) and LH ($R_L$, $R_H$) cases [3,5].

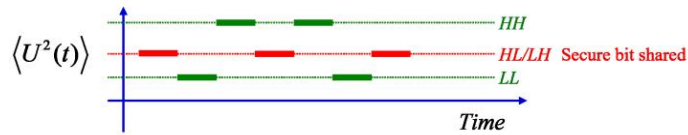

**Fig. 2.** Illustration of the mean-square voltage levels in the wire versus time during the operation of the key exchanger. The H and L indexes stand for the chosen resistors. The HL and LH levels are identical which makes the bit values corresponding to the HL and LH cases secure. On the other hand, the HH and LL levels are different, making them easily distinguishable by Eve. A similar graph exists for the mean-square currents [50-51,100].

The noise spectra $S_u(f)$ and $S_i(f)$ of the voltage $U(t)$ and current $I(t)$ in the wire, respectively, are given by the Johnson-Nyquist formulas of thermal noise [3,5]:

$$S_u(f) = 4kTR_p, \tag{1}$$

$$S_i(f) = \frac{4kT}{R_s}, \tag{2}$$

where $k$ is the Boltzmann constant, $T$ is the noise temperature and $R_p$ and $R_s$ are the parallel and serial resultant values of the connected resistors, respectively. In the LH and HL cases, the resultant values are:

$$R_{pHL} = R_{pLH} = \frac{R_H R_L}{R_H + R_L}, \tag{3}$$

$$R_{sHL} = R_{sLH} = R_H + R_L. \tag{4}$$

Leveraging Eqs. (1)-(4), it can be implied that the HL and LH resistor situations provide the same mean-square noise voltage spectra and noise current spectra, because both the parallel and serial resultant resistances are identical between the HL and LH states, respectively [50-51,100]. The above conditions can be summarized as:

$$U_{HL} = U_{LH}, \tag{5}$$

$$I_{HL} = I_{LH}, \tag{6}$$

$$P_{HL} = P_{LH} = 0, \tag{7}$$

where the voltage $U_{LH}$, $U_{HL}$, $I_{LH}$, and $I_{HL}$ effective (RMS) values stand for the voltage and current amplitudes in the wire, and $P$ is the mean power flow between Alice and Bob during the secure key exchange situation.

### *1.2. The Vadai, Mingesz and Gingl (VMG)-KLJN scheme*

Vadai, Mingesz and Gingl (VMG) introduced a genuine modification of the KLJN scheme [49] and proposed that, instead of using identical resistor pairs ($R_H$, $R_L$); it is still possible to maintain (*seemingly* [50-52]) perfectly secure communications, by using four *arbitrary* resistors: $R_{HA}$ and $R_{LA}$ at Alice, and $R_{HB}$ and $R_{LB}$ at Bob, (see Figure 3). As opposed to the ideal KLJN scheme, the VMG-KLJN key exchanger system is not in thermal equilibrium, as VMG showed that the four arbitrary resistors require different noise temperatures to guarantee that the voltage and current spectra and mean-square/effective values, and the net power flow in the wire are identical for the HL and LH cases [49].

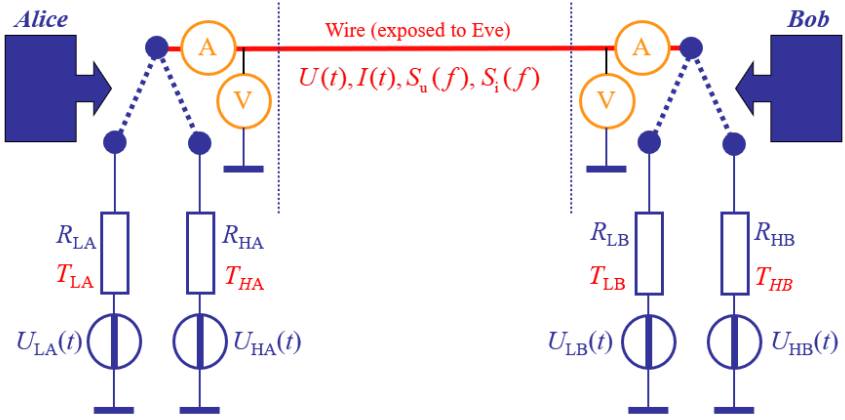

**Fig. 3.** The core of the VMG-KLJN secure key exchanger scheme [49-52]. The four resistors are different and can be freely chosen (with some limitations). The voltage generators $U_{HA}(t)$, $U_{LA}(t)$, $U_{HB}(t)$ and $U_{LB}(t)$ represent the thermal noise of the resistors $R_{HA}$, $R_{LA}$, $R_{HB}$ and $R_{LB}$ respectively. $T_{HA}$, $T_{LA}$, $T_{HB}$ and $T_{LB}$ represent the noise temperature of these resistors, respectively. The temperature of one of the resistors is freely chosen, and the other three temperatures depend on the corresponding resistor values and are given by the VMG Equations (9-11) [49].

The voltage generators $U_{HA}(t)$, $U_{LA}(t)$, $U_{HB}(t)$ and $U_{LB}(t)$ represent the thermal noise of the resistors $R_{HA}$, $R_{LA}$, $R_{HB}$ and $R_{LB}$ respectively. $T_{HA}$, $T_{LA}$, $T_{HB}$ and $T_{LB}$ represent the noise temperature of these resistors, respectively.

Equation 7 is replaced by the one below, indicating that non-zero power flow is possible:

$$P_{LH} = P_{HL} . \tag{8}$$

The temperature of one of the resistors is freely chosen, and the other three temperatures depend on the corresponding resistor values and are given by the VMG Equations (9-11) [49]:

$$U_{HB}^2 = U_{LA}^2 \frac{R_{LB}(R_{HA} + R_{HB}) - R_{HA} R_{HB} - R_{HB}^2}{R_{LA}^2 + R_{LB}(R_{LA} - R_{HA}) - R_{HA} R_{LA}} = 4kT_{HB} R_{HB} B \quad , \tag{9}$$

$$U_{HA}^2 = U_{LA}^2 \frac{R_{LB}(R_{HA} + R_{HB}) + R_{HA}R_{HB} + R_{HA}^2}{R_{LA}^2 + R_{LB}(R_{LA} + R_{HB}) + R_{HB}R_{LA}} = 4kT_{HA}R_{HA}B \tag{10}$$

and

$$U_{LB}^2 = U_{LA}^2 \frac{R_{LB}(R_{HA} - R_{HB}) - R_{HA}R_{HB} + R_{LB}^2}{R_{LA}^2 + R_{LA}(R_{HB} - R_{HA}) - R_{HA}R_{HB}} = 4kT_{LB}R_{LB}B \ . \tag{11}$$

### *1.3. Transient Attack against the KLJN scheme*

The d'Alembert equation describes propagating lossless fluctuations—which may or may not be waves—in a linear medium to model the propagation of voltage in the cable [90-92]:

$$U(t,x) = U_{+}\left(t - \frac{x}{v}\right) + U_{-}\left(t + \frac{x}{v}\right), \tag{12}$$

where $U_{+}$ and $U_{-}$ are voltage components of waves propagating to the right and left along the *x* axis, respectively; and *v* is propagation velocity. Gunn, et al [90] tried to use this delay picture to attack the steady-state KLJN scheme but the approach turned out to be flawed and was refuted [91,92]. On the other hand, Gunn, et al, [47] proposed a transient attack based on Equation (12). A generic defense against transient attacks was published earlier [46], however that has never been tested due to the complexity of its realization. The transient attack proposed in [47] was addressed recently [100] and a successful defense protocol was proposed and demonstrated, see the description below.

Transients start at the beginning of the bit exchange period when Alice and Bob connect the wire to their chosen resistor and the voltage and current in the wire abruptly change [100]. A transient attack [46-47,53] takes place during the short duration of time before the noise voltage and current in the cable reach thermal equilibrium, which is during when the mixing and equilibration of noise propagation in the cable takes place [5,100]. Eve can perform passive measurements using the transients to extract secure information, because the transients and reflections depend on the value of the terminating resistance of the cable, which are different between Alice and Bob [100].

Earlier, another defense scheme against the transient attack [47] was published without experimental verification [53]. Recently, inspired by [53], an advanced defense protocol [100] was proposed and explored by computer simulations.

Specific voltage and current based transient attacks [47,100] are covered in the present paper. During the transient period, the mean square noise voltages and currents are measured at Alice's and Bob's terminals. Eve uses comparison for the attack protocol: Eve decides about the value of the terminating resistor ($R_H$ or $R_L$) by comparing the mean square voltages (or currents), respectively [100]. During the transient period, the higher mean-square value indicates the lower terminating resistance (note, there is a typo about this fact in the text of [100]).

## 2. The formerly demonstrated, most efficient defense method

The previously demonstrated defense protocol [100] has revealed the most efficient approach. This protocol, as delineated below, necessitates the utilization of external noise generators with an augmented effective temperature, effectively overpowering the inherent thermal noise of the resistors. These external noise sources are pre-generated and stored in databases. From these stored noise databases, the amplitude and slope of the initial voltage amplitude are meticulously selected to meet the following criteria:

i. Ensuring that the instantaneous values of the driving noise voltages at both Alice's and Bob's terminals start at zero, thus mitigating the impact of transient reflections.

ii. Maintaining the starting slope of these noise segments at a predetermined ratio, thus guaranteeing the initiation of uniform charging currents and voltage levels:

$$m_{HL} = \frac{m_{HA}}{m_{LB}} = \frac{R_{HA} + Z_0}{R_{LB} + Z_0}, \tag{13}$$

or,

$$m_{LH} = \frac{m_{HB}}{m_{LA}} = \frac{R_{HB} + Z_0}{R_{LA} + Z_0}, \tag{14}$$

where, $m_{HL}$ and $m_{LH}$ are the required ratios of the slopes $m_{HA}$ and $m_{LB}$ and that of $m_{HB}$ and $m_{LA}$, respectively. $Z_0$ is the cable characteristic impedance. The H and L indexes refer to the $R_H$ and $R_L$ resistors, respectively. Note, it is assumed that the situation is the typical one, $Z_0 << [R_{LA,} R_{LB}]$.

## 3. Exploring the transient attack against the VMG-KLJN scheme

Details of the LTSPICE cable simulator setup and the process of generating the anti-aliased Gaussian white noise has been described in our former paper [100].

Both the HL and LH states are covered by the cable simulations. For the HL state, the resistance values are $R_{HA}$ = 11 kΩ, $R_{LB}$ = 2 kΩ; and for the LH state, the resistance values are $R_{LA}$ = 3 kΩ, $R_{HB}$ = 9 kΩ; the cable characteristic impedance is $Z_0$ = 50 Ω; the noise bandwidth is $B$ = 5 kHz. The cable length is $L$ = 2000 m that yields a cable fly time $t_f = 10^{-5}$ s.

Table 1 shows the noise temperatures of the resistors used to demonstrate the transient attack. For the temperature data, $U_{LA}$ = 1V is used.

| HL | | LH | | Noise Temperature | | | |
|---|---|---|---|---|---|---|---|
| $R_{HA}$ kΩ | $R_{LB}$ kΩ | $R_{LA}$ kΩ | $R_{HB}$ kΩ | $T_{HA}$ (K) | $T_{LB}$ (K) | $T_{LA}$ (K) | $T_{HB}$ (K) |
| 11 | 2 | 3 | 9 | 1.43x10$^{15}$ | 1.72x10$^{15}$ | 1.21x10$^{15}$ | 1.41x10$^{15}$ |

**Table-1.** Noise temperatures of the resistors used in the HL and LH states. For the temperature data, noise bandwidth $B$ = 5 kHz and $U_{LA}$ = 1V are used. All the four resistors have different noise temperatures.

The duration of the simulation is varied across 1 to 4 times the fly time of $\tau = 10^{-5}$ s and $4 \times 10^{-5}$ s, respectively. Thousand runs with independent noises are repeated for each scenario to calculate the probability $p_E$ of Eve's successful guessing of the bit value. The whole process to determine the value of $p_E$ (for both voltage and current transients) is repeated 10 times independently. The mean and the standard deviation are shown below.

### 3.1. Simulation results for the HL and the LH case

In order to establish a baseline, first the *no-defense* case was simulated, for both the HL and LH states, where the driving noises of the bit exchange are started abruptly at random and independent voltage amplitudes at the two ends. Simulation was being performed over two different transient observation time $\tau = 10^{-5}$ s and $4 \times 10^{-5}$ s for both the HL and LH cases. All the four cases (case A and B for the HL state, and case C and D for the LH state) indicate a huge information leak with $p_{E,V}$ and $p_{E,I}$ (the $p_E$ values for voltage and current transients, respectively) being further away from 0.5, see Table 2.

| Case | $\tau$ (10$^{-5}$s) | HL vs LH State | Resistors in the Loop | $p_{E,V}$ | $p_{E,I}$ |
|---|---|---|---|---|---|
| A | 1 | HL | $R_{HA}$=11kΩ, $R_{LB}$=2kΩ | 0.664 ± 0.012 | 0.664 ± 0.012 |
| B | 4 | HL | $R_{HA}$=11kΩ, $R_{LB}$=2kΩ | 0.756 ± 0.013 | 0.786 ± 0.009 |
| C | 1 | LH | $R_{LA}=3\text{k}\Omega, R_{HB}=9\text{k}\Omega$ | 0.640 ± 0.014 | 0.640 ± 0.014 |
| D | 4 | LH | $R_{LA}=3\text{k}\Omega, R_{HB}=9\text{k}\Omega$ | 0.636 ± 0.018 | 0.652 ± 0.016 |

**Table 2.** Demonstration of the mean-square voltage-based (or mean-square current-based) transient attack, with no defense being applied, over two different transient observation time ($\tau$) for both the HL and LH cases. Case (A) and (C) represent transient observation times of $\tau = 10^{-5}$ s; and case (B) and (D) represent transient observation time of $\tau = 4 \times 10^{-5}$ s, for the HL and LH cases, respectively. The parameters of the simulation: $T_{HA}$ = 1.43x10$^{15}$ K, $T_{LB}$ =1.72x10$^{15}$ K, $T_{LA}$ = 1.21x10$^{15}$ K, $T_{HB}$ =1.41x10$^{15}$ K, $B$ = 5 kHz, $Z_0$ = 50 Ω. $p_{E,V}$ and $p_{E,I}$ are Eve's probability of correctly guessing the bit values with voltage and current based transient attacks, respectively.

The most efficient defense against transient attack [100] is that Alice's and Bob's generator noises start from zero voltage at the beginning of the BEP, and the starting slopes are scaled at the ratio of $m_{HL}$ or $m_{LH}$, see Equation (13)-(14), for the HL or LH state, respectively. For all the four cases (case E and F for the HL state, and case G and H for the LH state), with the defense being applied, Eve does not have sufficient information to crack the VMG-KLJN scheme, as $p_{E,V}$ and $p_{E,I}$ are close to 0.5, indicating near to perfect security, see Table 3.

| Case | $\tau$ (10$^{-5}$s) | HL vs LH State | Resistors in the Loop | $p_{E,V}$ | $p_{E,I}$ |
|---|---|---|---|---|---|
| E | 1 | HL | $R_{HA}=11\text{k}\Omega, R_{LB}=2\text{k}\Omega$ | 0.502 ± 0.016 | 0.502 ± 0.016 |
| F | 4 | HL | $R_{HA}=11\text{k}\Omega, R_{LB}=2\text{k}\Omega$ | 0.535 ± 0.013 | 0.534 ± 0.015 |
| G | 1 | LH | $R_{LA}=3\text{k}\Omega, R_{HB}=9\text{k}\Omega$ | 0.504 ± 0.014 | 0.504 ± 0.014 |
| H | 4 | LH | $R_{LA}=3\text{k}\Omega, R_{HB}=9\text{k}\Omega$ | 0.522 ± 0.023 | 0.520 ± 0.018 |

**TABLE 3.** Demonstration of the most efficient defense protocol against the mean-square voltage-based (or mean-square current-based) transient attack; Alice and Bob's noise generators are started from zero voltage and with matching slope at the ratio of $m_{HL}$ or $m_{LH}$, see Equation (13)-(14), for the HL and LH case, respectively. Case (E) and (G) represent transient observation times of $\tau = 10^{-5}$ s; and case (F) and (H) represent transient observation time of $\tau = 4 \times 10^{-5}$ s, for the HL and LH cases, respectively. The parameters of the simulation: $T_{HA}$ = 1.43x10$^{15}$ K, $T_{LB}$ =1.72x10$^{15}$ K, $T_{LA}$ = 1.21x10$^{15}$ K, $T_{HB}$ =1.41x10$^{15}$ K, $B$ = 5 kHz, $Z_0$ = 50 Ω. The probabilities $p_{E,V}$ and $p_{E,I}$ of correctly guessing the bit values are very close to the value 0.5 (perfect security), thus the defense works efficiently.

## 4. Conclusion

Simulation results of cable dynamics have demonstrated the vulnerability of the VMG-KLJN scheme to transient attacks. However, as delineated in reference [100], an effective defense strategy can be implemented. This strategy entails the meticulous selection of initial voltage amplitudes and slopes for the two noise generators, ultimately fortifying the system against transient attacks. Notably, this defense approach proves equally effective when applied to both the ideal KLJN and the VMG-KLJN secure key exchange methodologies.

## References

[1]  C.E. Shannon, Communication theory of secrecy systems, *Bell Systems Technical Journal* **28** (1949) 656–715.
[2]  Y. Liang, H.V. Poor and S. Shamai, Information theoretic security, *Foundations Trends Commun. Inform. Theory* **5** (2008) 355–580.
[3]  L.B. Kish, Totally secure classical communication utilizing Johnson (-like) noise and Kirchhoff's law, *Phys. Lett. A* **352** (2006) 178-182.
[4]  A. Cho, simple noise may stymie spies without quantum weirdness, *Science* **309** (2005) 2148-2148.
[5]  L.B. Kish, The Kish Cypher: The Story of KLJN for Unconditional Security, New Jersey: World Scientific, (2017).
[6]  L.B. Kish and C.G. Granqvist, On the security of the Kirchhoff-law–Johnson-noise (KLJN) communicator, *Quant. Inform. Proc.* **13** (2014) (10) 2213-2219.
[7]  L.B. Kish, D. Abbott, and C. G. Granqvist, Critical analysis of the Bennett–Riedel attack on secure cryptographic key distributions via the Kirchhoff-law–Johnson-noise scheme, *PloS One* **8** (2013) e81810.
[8]  H. P. Yuen, Security of quantum key distribution, *IEEE Access* **4** (2016) 7403842.
[9]  C. H. Bennett and G. Brassard, Quantum Cryptography: Public Key Distribution and Coin Tossing, Proc. IEEE Int. Conf. Comp., Syst., Signal 277, Process. 1, 175–179 (1984).
[10]  S. Sajeed, A. Huang, S. Sun, F. Xu, V. Makarov, and M. Curty, Insecurity of detector-device-independent quantum key distribution, *Phys. Rev. Lett.* **117** (2016) 250505.
[11]  H. P. Yuen, Essential elements lacking in security proofs for quantum key distribution, *Proc. SPIE* **8899** (2013) 88990J.
[12]  O. Hirota, Incompleteness and limit of quantum key distribution theory. Available online arXiv:1208.2106.
[13]  N. Jain, E. Anisimova, I. Khan, V. Makarov, C. Marquardt and G. Leuchs, Trojan-horse attacks threaten the security of practical quantum cryptography, *New J. Phys.* **16** (2014) 123030.
[14]  I. Gerhardt, Q. Liu, A. Lamas-Linares, J. Skaar, C. Kurtsiefer and V. Makarov, Full-field implementation of a perfect eavesdropper on a quantum cryptography system, *Nature Commun.* **349** (2012)
[15]  L. Lydersen, C. Wiechers, C. Wittmann, D. Elser, J. Skaar and V. Makarov, Hacking commercial quantum cryptography systems by tailored bright illumination, *Nature*

*Photon.* **4** (2010) 686–689.

[16] I. Gerhardt, Q. Liu, A. Lamas-Linares, J. Skaar, V. Scarani, V. Makarov and C. Kurtsiefer, Experimentally faking the violation of Bell's inequalities, *Phys. Rev. Lett.* **107** (2011) 170404.

[17] V. Makarov and J. Skaar, Fakes states attack using detector efficiency mismatch on SARG04, phase-time, DPSK, and Ekert protocols, *Quant. Inform. Comput.* **8** (2008) 622–635.

[18] C. Wiechers, L. Lydersen, C. Wittmann, D. Elser, J. Skaar, C. Marquardt, V. Makarov and G. Leuchs, After-gate attack on a quantum cryptosystem, *New J. Phys.* **13** (2011) 013043.

[19] L. Lydersen, C. Wiechers, C. Wittmann, D. Elser, J. Skaar and V. Makarov, Thermal blinding of gated detectors in quantum cryptography, *Opt. Express* **18** (2010) 27938–27954.

[20] N. Jain, C. Wittmann, L. Lydersen, C. Wiechers, D. Elser, C. Marquardt, V. Makarov and G. Leuchs, Device calibration impacts security of quantum key distribution, *Phys. Rev. Lett.* **107** (2011) 110501.

[21] L. Lydersen, J. Skaar and V. Makarov, Tailored bright illumination attack on distributed-phase-reference protocols, *J. Mod. Opt.* **58** (2011) 680–685.

[22] L. Lydersen, M. K. Akhlaghi, A. H. Majedi, J. Skaar and V. Makarov, Controlling a superconducting nanowire single-photon detector using tailored bright illumination, *New J. Phys.* **13** (2011) 113042.

[23] L. Lydersen, V. Makarov and J. Skaar, Comment on "Resilience of gated avalanche photodiodes against bright illumination attacks in quantum cryptography", *Appl. Phys. Lett.* **98** (2011) 231104.

[24] P. Chaiwongkhot, K.B. Kuntz, Y. Zhang, A. Huang, J.P. Bourgoin, S. Sajeed, N. Lütkenhaus, T. Jennewein and V. Makarov, Eavesdropper's ability to attack a free-space quantum-key-distribution receiver in atmospheric turbulence, *Phys. Rev. A* **99** (2019) 062315.

[25] G. Gras, N. Sultana, A. Huang, T. Jennewein, F. Bussières, V. Makarov, and H. Zbinden, Optical control of single-photon negative-feedback avalanche diode detector, *J. Appl. Phys.* **127** (2020) 094502.

[26] A. Huang, R. Li, V. Egorov, S. Tchouragoulov, K. Kumar, and V. Makarov, Laser-damage attack against optical attenuators in quantum key distribution, *Phys. Rev. Appl.* **13** (2020) 034017.

[27] A. Huang, Á. Navarrete, S.-H. Sun, P. Chaiwongkhot, M. Curty, and V. Makarov, Laser-seeding attack in quantum key distribution, *Phys. Rev. Appl.* **12** (2019) 064043.

[28] V. Chistiakov, A. Huang, V. Egorov, and V. Makarov, Controlling single-photon detector ID210 with bright light, *Opt. Express* **27** (2019) 32253.

[29] A. Fedorov, I. Gerhardt, A. Huang, J. Jogenfors, Y. Kurochkin, A. Lamas-Linares, J.-Å. Larsson, G. Leuchs, L. Lydersen, V. Makarov, and J. Skaar, Comment on "Inherent security of phase coding quantum key distribution systems against detector blinding attacks" (2018 Laser Phys. Lett. 15 095203), *Laser Phys. Lett.* **16** (2019) 019401.

[30] A. Huang, S. Barz, E. Andersson and V. Makarov, Implementation vulnerabilities in

general quantum cryptography, *New J. Phys.* **20** (2018) 103016.

[31] P.V.P. Pinheiro, P. Chaiwongkhot, S. Sajeed, R. T. Horn, J.-P. Bourgoin, T. Jennewein, N. Lütkenhaus and V. Makarov, Eavesdropping and countermeasures for backflash side channel in quantum cryptography, *Opt. Express* **26** (2018) 21020.

[32] A. Huang, S.-H. Sun, Z. Liu and V. Makarov, Quantum key distribution with distinguishable decoy states, *Phys. Rev. A* **98** (2018) 012330.

[33] H. Qin, R. Kumar, V. Makarov and R. Alléaume, Homodyne-detector-blinding attack in continuous-variable quantum key distribution, *Phys. Rev. A* **98** (2018) 012312.

[34] S. Sajeed, C. Minshull, N. Jain and V. Makarov, Invisible Trojan-horse attack, *Sci. Rep.* **7** (2017) 8403.

[35] P. Chaiwongkhot, S. Sajeed, L. Lydersen and V. Makarov, Finite-key-size effect in commercial plug-and-play QKD system, *Quantum Sci. Technol.* **2** (2017) 044003.

[36] A. Huang, S. Sajeed, P. Chaiwongkhot, M. Soucarros, M. Legré and V. Makarov, Testing random-detector-efficiency countermeasure in a commercial system reveals a breakable unrealistic assumption, *IEEE J. Quantum Electron*. **52** (2016) 8000211.

[37] V. Makarov, J.-P. Bourgoin, P. Chaiwongkhot, M. Gagné, T. Jennewein, S. Kaiser, R. Kashyap, M. Legré, C. Minshull and S. Sajeed, Creation of backdoors in quantum communications via laser damage, *Phys. Rev. A* **94** (2016) 030302.

[38] S. Sajeed, P. Chaiwongkhot, J.-P. Bourgoin, T. Jennewein, N. Lütkenhaus and V. Makarov, Security loophole in free-space quantum key distribution due to spatial-mode detector-efficiency mismatch, *Phys. Rev. A* **91** (2015) 062301.

[39] N. Jain, B. Stiller, I. Khan, V. Makarov, Ch. Marquardt and G. Leuchs, Risk analysis of Trojan-horse attacks on practical quantum key distribution systems, *IEEE J. Sel. Top. Quantum Electron*. **21** (2015) 6600710.

[40] M.G. Tanner, V. Makarov and R. H. Hadfield, Optimised quantum hacking of superconducting nanowire single-photon detectors, *Opt. Express* **22** (2014) 6734.

[41] A.N. Bugge, S. Sauge, A. M. M. Ghazali, J. Skaar, L. Lydersen and V. Makarov, Laser damage helps the eavesdropper in quantum cryptography, *Phys. Rev. Lett.* **112** (2014) 070503.

[42] Q. Liu, A. Lamas-Linares, C. Kurtsiefer, J. Skaar, V. Makarov and I. Gerhardt, A universal setup for active control of a single-photon detector, *Rev. Sci. Instrum.* **85** (2014) 013108.

[43] See https://www.nsa.gov/Cybersecurity/Quantum-Key-Distribution-QKD-and-Quantum-Cryptography-QC/ for information about why NSA does not support the usage of QKD or QC to protect communications

[44] S. Sajeed, I. Radchenko, S. Kaiser, J.-P. Bourgoin, A. Pappa, L. Monat, M. Legré and V. Makarov, Attacks exploiting deviation of mean photon number in quantum key distribution and coin tossing, *Phys. Rev. A* **91** (2015) 032326.

[45] R. Mingesz, Z. Gingl and L.B. Kish, Johnson(-like)-noise-Kirchhoff-loop based secure classical communicator characteristics, for ranges of two to two thousand kilometers, via model-line, *Phys. Lett. A* **372** (2008) 978–984.

[46] L.B. Kish, Enhanced secure key exchange systems based on the Johnson-noise scheme, *Metrol. Meas. Syst.* **20** (2013) 191-204.

[47] L.J. Gunn, A. Allison and D. Abbott, A new transient attack on the Kish key
12

distribution system, *IEEE Access* **3** (2015) 1640-1648.
[48] G. Vadai, Z. Gingl and R. Mingesz, Generalized attack protection in the Kirchhoff-law-Johnson-noise key exchanger, *IEEE Access*, **4** (2016) 1141-1147.
[49] G. Vadai, R. Mingesz and Z. Gingl, Generalized Kirchhoff-law-Johnson-noise (KLJN) secure key exchange system using arbitrary resistors, *Scientific reports* **5** (2015) 13653.
[50] S. Ferdous, C. Chamon and L.B. Kish, Comments on the "Generalized" KJLN Key Exchanger with Arbitrary Resistors: Power, Impedance, Security, *Fluctuation and Noise Lett.* **20**, No. 01, 2130002 (2021).
[51] S. Ferdous, C. Chamon and L. B. Kish, Current Injection and Voltage Insertion Attacks Against the VMG-KLJN Secure Key Exchanger, *Fluctuation Noise Lett.* **22**, 2350009 (2023).
[52] C. Chamon, S. Ferdous and L.B. Kish, Random number generator attack against the Kirchhoff-law-Johnson-noise secure key exchange protocol, *Fluctuation Noise Lett.* **21**, 2250027 (2022).
[53] L.B. Kish and C. G. Granqvist, Comments on "A new transient attack on the Kish key distribution system", *Metrol. Meas. Syst.* **23** (2015) 321-331.
[54] L.B. Kish and C.G. Granqvist, Random-resistor-random-temperature Kirchhoff-law-Johnson-noise(RRRT -KLJN) key exchange, *Metrol. Meas. Syst.* **23** (2016) 3-11.
[55] L.B. Kish and T. Horvath, Notes on recent approaches concerning the Kirchhoff-law–Johnson-noise based secure key exchange, *Phys. Lett. A* **373** (2009) 2858-2868.
[56] J. Smulko, Performance analysis of the 'intelligent Kirchhoff-law–Johnson-noise secure key exchange", *Fluct. Noise Lett.* **13** (2014) 1450024.
[57] R. Mingesz, L.B. Kish , Z. Gingl, C.G. Granqvist, H. Wen, F. Peper, T. Eubanks and G. Schmera, Unconditional security by the laws of classical physics, *Metrol. Meas. Syst.* **XX** (2013) 3–16.
[58] T. Horvath, L.B. Kish and J. Scheuer, Effective privacy amplification for secure classical communications, *EPL* **94** (2011), 28002.
[59] Y. Saez and L.B. Kish, Errors and their mitigation at the Kirchhoff-law-Johnson-noise secure key exchange, *PLoS ONE* **8** (2013) e81103.
[60] R. Mingesz, G. Vadai and Z. Gingl, What kind of noise guarantees security for the Kirchhoff-Loop-Johnson-Noise key exchange?, *Fluct. Noise Lett.* **13** (2014) 1450021.
[61] Y. Saez, L.B. Kish, R. Mingesz, Z. Gingl and C.G. Granqvist, Current and voltage based bit errors and their combined mitigation for the Kirchhoff-law-Johnson-noise secure key exchange, *J. Comput. Electron.* **13** (2014) 271–277.
[62] Y. Saez, L.B. Kish, R. Mingesz, Z. Gingl and C.G. Granqvist, Bit errors in the Kirchhoff-law-Johnson-noise secure key exchange, *Int. J. Mod. Phys.*: Conference Series **33** (2014) 1460367.
[63] Z. Gingl and R. Mingesz, Noise properties in the ideal Kirchhoff-Law-Johnson-Noise secure communication system, *PLoS ONE* **9** (2014) e96109.
[64] L.B. Kish and R. Mingesz, Totally secure classical networks with multipoint telecloning (teleportation) of classical bits through loops with Johnson-like noise, *Fluct. Noise Lett.* **6** (2006) C9–C21.

[65] L.B. Kish, Methods of using existing wire lines (power lines, phone lines, internet lines) for totally secure classical communication utilizing Kirchoff's Law and Johnson-like noise, (2006), Available online https://arXiv.org/abs/physics/0610014.

[66] L.B. Kish and F. Peper, Information networks secured by the laws of physics, *IEICE Trans. Fund. Commun. Electron. Inform. Syst.* **E95–B5** (2012) 1501– 1507.

[67] E. Gonzalez, L.B. Kish, R.S. Balog and P. Enjeti, Information theoretically secure, enhanced Johnson noise based key distribution over the smart grid with switched filters, *PloS One* **8** (2013) e70206.

[68] E. Gonzalez, L.B. Kish and R. Balog, Encryption Key Distribution System and Method, U.S. Patent # US9270448B2 (granted 2/2016), https://patents.google.com/patent/US9270448B2.

[69] E. Gonzalez, R. Balog, R. Mingesz and L.B. Kish, Unconditionally security for the smart power grids and star networks, *23rd International Conference on Noise and Fluctuations (ICNF 2015),* Xian, China, June 2-5, 2015.

[70] E. Gonzalez, R. S. Balog and L.B. Kish, Resource requirements and speed versus geometry of unconditionally secure physical key exchanges, *Entropy* **17** (2015) 2010–2014.

[71] E. Gonzalez and L.B. Kish, "Key Exchange Trust Evaluation in Peer-to-Peer Sensor Networks With Unconditionally Secure Key Exchange", *Fluct. Noise Lett.* **15** (2016) 1650008.

[72] L.B. Kish and O. Saidi, Unconditionally secure computers, algorithms and hardware, such as memories, processors, keyboards, flash and hard drives, *Fluct. Noise Lett.* **8** (2008) L95–L98.

[73] L.B. Kish, K. Entesari, C.-G. Granqvist and C. Kwan, Unconditionally secure credit/debit card chip scheme and physical unclonable function, *Fluct. Noise Lett.* **16** (2017) 1750002.

[74] L.B. Kish and C. Kwan, Physical unclonable function hardware keys utilizing Kirchhoff-law-Johnson noise secure key exchange and noise-based logic, *Fluct. Noise Lett.* **12** (2013) 1350018.

[75] Y. Saez, X. Cao, L.B. Kish and G. Pesti, Securing vehicle communication systems by the KLJN key exchange protocol, *Fluct. Noise Lett.* **13** (2014) 1450020.

[76] X. Cao, Y. Saez, G. Pesti and L.B. Kish, On KLJN-based secure key distribution in vehicular communication networks, *Fluct. Noise Lett.* **14** (2015) 1550008.

[77] L.B. Kish and C. G. Granqvist, Enhanced usage of keys obtained by physical, unconditionally secure distributions, *Fluct. Noise Lett.* **14** (2015) 1550007.

[78] L.B. Kish, Protection against the man-in-the-middle-attack for the Kirchhoff-loop-Johnson (-like)-Noise Cipher and Expansion by Voltage-Based Security, *Fluct. Noise Lett.* **6** (2006) L57-L63.

[79] H.P. Chen, M. Mohammad and L.B. Kish, Current injection attack against the KLJN secure key exchange, *Metrol. Meas. Syst.* **23** (2016) 173-181.

[80] M.Y. Melhem and L.B. Kish, Generalized DC loop current attack against the KLJN secure key exchange scheme, *Metrol. Meas. Syst.* **26** (2019) 607-616.

[81] M.Y. Melhem and L.B. Kish, A static-loop-current attack against the Kirchhoff-law-

Johnson-noise (KLJN) secure key exchange system, *Applied Sciences* **9** (2019) 666.
[82] M.Y. Melhem and L.B. Kish, The problem of information leak due to parasitic loop currents and voltages in the KLJN secure key exchange scheme, *Metrol. Meas. Syst.* **26** (2019) 37–40.
[83] F. Hao, Kish's key exchange scheme is insecure, *IEE Proceedings-Information Security*, **153** (2006) 141-142.
[84] L.B. Kish, Response to Feng Hao's paper "Kish's key exchange scheme is insecure", *Fluct. Noise Lett.* **6** (2006) C37-C41.
[85] L.B. Kish and J. Scheuer, Noise in the wire: the real impact of wire resistance for the Johnson (-like) noise based secure communicator, *Phys. Lett. A* **374** (2010) 2140-2142.
[86] L.B. Kish and C.-G. Granqvist, Elimination of a second-law-attack, and all cable-resistance-based attacks, in the Kirchhoff-law-Johnson-noise (KLJN) secure key exchange system, *Entropy*, **16** (2014) 5223-5231.
[87] M.Y. Melhem, C. Chamon, S. Ferdous and L.B. Kish, Alternating (AC) Loop Current Attacks Against the KLJN Secure Key Exchange Scheme, Fluctuation and Noise Letters, Vol. **20**, No. 3 (2021) 2150050, doi: 10.1142/S0219477521500504.
[88] H.-P. Chen, E. Gonzalez, Y. Saez and L.B. Kish, Cable capacitance attack against the KLJN secure key exchange, Information, **6** (2015) 719-732.
[89] M.Y. Melhem and L.B. Kish, Man in the middle and current injection attacks against the KLJN key exchanger compromised by DC sources, Fluctuation and Noise Letters, Vol. **20**, No. 2 (2021) 2150011, DOI: 10.1142/ S0219477521500115.
[90] L.J. Gunn, A. Allison and D. Abbott, A directional wave measurement attack against the Kish key distribution system, *Scientific Reports* **4** (2014) 6461.
[91] H.-P. Chen, L.B. Kish and C. G. Granqvist, On the "Cracking" Scheme in the Paper "A Directional Coupler attack against the Kish key distribution system" by Gunn, Allison and Abbott, *Metrol. and Meas. Syst.* **21** (2014) 389-400.
[92] H.-P. Chen, L.B. Kish, C.-G. Granqvist, and G. Schmera, Do electromagnetic waves exist in a short cable at low frequencies? What does physics say?, *Fluct. Noise Lett.* **13** (2014) 1450016.
[93] L.B. Kish, Z. Gingl, R. Mingesz, G. Vadai, J. Smulko and C.-G. Granqvist, Analysis of an Attenuator artifact in an experimental attack by Gunn–Allison–Abbott against the Kirchhoff-law–Johnson-noise (KLJN) secure key exchange system, *Fluct. Noise Lett.* **14** (2015) 1550011.
[94] P.L. Liu, A Complete Circuit Model for the Key Distribution System Using Resistors and Noise Sources, *Fluct. Noise Lett.* **19** (2020) 2050012.
[95] P.L. Liu, Re-Examination of the Cable Capacitance in the Key Distribution System Using Resistors and Noise Sources, *Fluct. Noise Lett.* **16** (2017) 1750025.
[96] P.L. Liu, A key agreement protocol using band-limited random signals and feedback, *IEEE J. of Lightwave Tech.*, **27** (2009) 5230-5234.
[97] C. Chamon and L.B. Kish, Perspective—On the thermodynamics of perfect unconditional security, Appl. Phys. Lett. **119**, 010501 (2021).
[98] C. Chamon, S. Ferdous and L.B. Kish, Deterministic Random Number Generator Attack against the Kirchhoff-Law-Johnson-Noise Secure Key Exchange Protocol,

Fluctuation and Noise Letters, Vol. **20**, No. 5 (2021) 2150046, DOI: 10.1142/S0219477521500462.

[99] L.B. Kish, Time synchronization protocol for the KLJN secure key exchange scheme, *Fluctuation and Noise Lett.* **21**, 2250046 (2022).

[100] S. Ferdous and L.B. Kish, Transient attacks against the Kirchhoff–Law–Johnson–Noise (KLJN) secure key exchanger, *Appl. Phys. Lett.* **122**, 143503 (2023).